\documentclass[3p,12pt]{elsarticle}

\usepackage{xcolor}
\usepackage{float}
\usepackage{amsmath}
\usepackage{etoolbox}
\usepackage{xcolor}
\usepackage{graphicx}
\usepackage{dcolumn}
\usepackage{bm}
\usepackage{multirow}
\usepackage{lineno,hyperref}

\usepackage{amssymb}

\usepackage[figuresright]{rotating}

\begin{document}

\begin{frontmatter}




\title{Modelling Growth, Remodelling and Damage of a Thick-walled Fibre-reinforced Artery with Active Response: Application to Cerebral Vasospasm and Treatment}


\author[CSL]{G. Pederzani }
\cortext[corauth]{Corresponding author}
\author[MECH,INS]{A. Grytsan}
\author[CSL]{A.G. Hoekstra}
\author[PITT]{A.M. Robertson}
\author[COMP,INS,PITT]{P.N. Watton \corref{corauth}}
\ead{ p.watton@sheffield.ac.uk }
\address[CSL]{Computational Science Lab, Instituut voor Informatica, Universiteit van Amsterdam, Netherlands}
\address[MECH]{Department of Mechanical Engineering, University of Sheffield, UK}
\address[COMP]{Department of Computer Science, University of Sheffield, UK}
\address[INS]{INSIGNEO Institute of \textit{in silico} Medicine, University of Sheffield, UK}
\address[PITT]{Department of Mechanical Engineering and Materials Science, University of Pittsburgh, USA}

\begin{abstract}
\textit{Background:}
Cerebral vasospasm, a prolonged constriction of cerebral arteries, is the first cause of morbidity and mortality for patients who survive hospitalisation after aneurysmal subarachnoid haemorrhage. Recently, it has been observed that (low pressure) stent-retrievers can be used to successfully treat the disease which has challenged a widely held viewpoint that damage to the extracellular matrix is necessary to mechanically resolve the constriction.  \\
\textit{Methods:}
We apply a 3D finite element rate-based constrained mixture model (rb-CMM) approach to simulate the artery, vasospasm, remodelling and treatment with stents. The artery is modelled as a thick-walled  fibre-reinforced constrained mixture subject to physiological pressure and axial stretch. The model accounts for distributions of collagen fibre homeostatic stretches, VSMC active response, remodelling and damage. After simulating vasospasm and subsequent remodelling of the artery to a new homeostatic state, we simulate treatment with commonly available stent-retrievers. Subsequently, we perform a parameter study to examine how arterial diameter and thickness affect the success of stent treatment.  \\
\textit{Results:}
The model predictions on the amount of pressure required to mechanically resolve the constriction are consistent with those applied by stent-retrievers. In agreement with reported clinical observations, our model predicts that stent-retrievers tend to be effective in arteries of up to about $3mm$ original diameter, but usually fail in larger ones. In addition, we observe that accounting for (physiological) variations in arterial wall thickness, the stent pressure requirement may need to double to successfully treat vasospasm. 
\textit{Conclusions:}
We have developed a novel rb-CMM that accounts for VSMC active response, remodelling and damage. Consistent with clinical observations, simulations predict that stent-retrievers can mechanically resolve vasospasm. Moreover, accounting for a patient's arterial properties is important for predicting likelihood of stent success. This \textit{in silico} tool has the potential to support clinical decision-making and to guide the development and evaluation of dedicated stents for personalised treatment of vasospasm. 
\end{abstract}

\begin{keyword}
cerebral vasospasm \sep finite element \sep mechanobiology \sep smooth muscle cell \sep stent 

\end{keyword}

\end{frontmatter}



\section{Introduction}
\label{sec:intro}

Cerebral vasospasm is a severe and prolonged constriction of a cerebral artery, most often caused by subarachnoid hemorrhage (SAH). It affects up to 70$\%$ of patients who survive hospitalisation after SAH and is the leading cause of mortality for this category of patients \cite{chalet_clinical_2023} due to the resulting reduction in blood flow and thus brain tissue oxygenation. Earliest detection occurs around day 3 post-SAH and, if pharmacological treatment is insufficient to resolve the contraction, mechanical treatment is required and  usually occurs around 7-10 days post-SAH.

Cerebral vasospasm is characterised by a complex, multifactorial pathophysiology where different factors play more or less important roles depending on the time elapsed from SAH. There is little doubt that the initial insult is caused by oxy-haemoglobin released by the extravasated blood \cite{Macdonald_Weir_1991, Findlay_Nisar_Darsaut_2016, Baggott_Aagaard-Kienitz_2014}. It initiates a chemical signalling cascade involving multiple pathways that, together with the release of inflammatory signals, results in an increase in vasoconstrictors and in scavengers of vasodilators (such as nitric oxide) thus causing a net increase in contraction of the vascular smooth muscle cells (VSMCs). This is consistent with the potential effectiveness of pharmacological treatment via vasodilators (such as nimodipine) especially in the first few days after SAH \cite{doi:10.1056/NEJM198303173081103, cochranereviewcalcium}. However the effectiveness of vasodilators decreases over time, together with arterial tissue compliance, while stiffness increases \cite{Yamaguchi-Okada_Nishizawa_Koide_Nonaka_2005, Macdonald1995, Matsui1994, Vorkapic1991a, Bevan1987}. This suggests a second phase of vasospasm, where some form of more irreversible remodelling occurs. The initial hypothesis has been that this remodelling consisted of increased collagen deposition in the arterial wall, which would explain the increased stiffness of the tissue, but while some histological studies found evidence of this \cite{Smith1985, Nagasawa1982}, other results were in disagreement with these findings \cite{Yamaguchi-Okada_Nishizawa_Koide_Nonaka_2005}. Some authors suggested another ECM protein could be responsible for the increased stiffness, but no direct evidence was found \cite{Yamaguchi-Okada_Nishizawa_Koide_Nonaka_2005}. 

The hypothesis of ECM remodelling in a second phase of vasospasm was challenged by the finding, first reported in 2017 by Bhogal et al. that stent-retrievers could be successful in treating vasospasm, at least in some cases \cite{Bhogal_Loh_Brouwer_Andersson_Söderman_2017, Bhogal_Paraskevopoulos_Makalanda_2017}. Indeed while balloon angioplasty, the most common treatment choice in case of ineffectiveness of pharmacological treatment, could apply sufficient pressure to damage the collagen matrix, stent-retrievers, which were originally designed for mechanical thrombectomy and typically apply pressures an order of magnitude lower than balloon angioplasty (30kPa vs. 300kPa), would not apply sufficient pressure to damage the ECM. Since then, a number of other groups have reported success in treating vasospasm with stent-retrievers \cite{Kwon_Lim_Koh_Park_Choi_Kim_Youm_Song_2019, Norby_Young_Siddiq_2019, Badger_Jankowitz_Shaikh_2020, Su_Ali_Pukenas_Favilla_Zanaty_Hasan_Kung_2020, Gupta_Woodward_2022, Hensler_Wodarg_Madjidyar_Peters_Cohrs_Jansen_Larsen_2022, Khanafer_Cimpoca_Bhogal_Bäzner_Ganslandt_Henkes_2022, Lopez-Rueda_2022}. We sought to investigate the mechanism of action of stent-assisted angioplasty by means of computational modelling.

Based on the theoretical idea of homeostasis and the hypothesis that cells act to maintain the tissue in a chemical, biological and mechanical homeostasis, our group formulated the hypothesis that before the ECM plays a significant role, it is remodelling of the vascular smooth muscle cells (VSMCs) that causes medically refractory vasospasm, and thus stent-retrievers could be sufficient to resolve the disease \cite{Bhogal2019}. Traditionally vasospasm has been thought of as a biphasic disease, with an acute phase characterised by increased contraction of VSMCs and a chronic phase characterised by some kind of ECM remodelling. Our hypothesis suggests the existence of an intermediate phase: after a first acute phase characterised by increased VSMC contraction, a second one is dominated by VSMC remodelling about the decreasing diameter; ECM remodelling occurs in a third phase. This new intermediate phase would explain the decreased effectiveness of vasodilators and increased stiffness of the wall, without involving the ECM. We also assume that this is the dominating factor in the time window during which patients typically show symptoms, about 7-10 days after SAH, although this may vary for different individuals. We also consider these phases to be largely overlapping, with continuous transitions from one dominating factor to the next.

The first mathematical model of vasospasm was proposed by Lodi and Ursino \cite{Lodi_Ursino_1999} with a focus on cerebral haemodynamics. They proposed a 0D model of the cerebral circulation with consideration of the collateral circulation, cerebrospinal fluid flow, intracranial pressure, venous haemodynamics and cerebral autoregulation. After replicating general patterns in cerebral haemodynamics observed in vasospasm patients, they performed a sensitivity analysis and in particular studied the reliability of transcranial Doppler measurement of blood flow velocity for the estimation of vasospasm severity. In  2007, Humphrey et al. and Baek et al. (\cite{Humphrey_Baek_Niklason_2007, Baek2007}) developed a 2D axisymmetric membrane model of cerebral vasospasm which coupled cerebral haemodynamics to the properties of the arterial wall (given by a constrained mixture of elastin, smooth muscle cells and collagen fibres), capturing the potential self-resolution of the condition. In 2019, motivated by clinically relevant questions regarding the mechanism of action of stent-assisted angioplasty in vasospasm cases, our group proposed the novel hypothesis on the role of VSMC remodelling mentioned earlier and implemented a membrane model to validate it \cite{Bhogal2019}. The consistency between the model predictions and clinical observations was an encouraging result, but the membrane formulation prevented us to capture heterogeneities in the geometry and structure of the arterial wall, as well as from explicitly simulating stent deployment and its effect on realistic irregular geometries.

In this work we propose a 3D finite element model of the development and treatment of vasospasm in a cerebral artery, which is to our knowledge the first of its kind. We use a rate-based constrained mixture approach with growth and remodelling, and model the development of vasospasm via increase of active stress and remodelling of vascular smooth muscle cells. We simulate mechanical treatment via the application of an additional pressure to the luminal surface and, based on a strain-based damage criterion, estimate the amount of pressure required to resolve the constriction. In Section \ref{sec:methods} we describe the mathematical model used in this work; in Section \ref{sec:results} we report model results in representing the artery in health, vasospasm and after mechanical treatment; in Section \ref{sec:discussion} we discuss the results and modelling assumptions, and finally in Section \ref{sec:conclusions} we summarise the contribution of this work.

\section{Methods}
\label{sec:methods}

The artery is modelled as a thick-walled fibre-reinforced cylinder subject to internal pressure and axial pre-stretch, which replicates \textit{in vivo} conditions. A rate-based constrained mixture (rb-CM) \cite{Watton2004} approach is adopted where the stress response of the tissue is given by the sum of the stress contributions of its microstructural constituents. For the purposes of this model we consider the more mechanically relevant constituents: elastin, vascular smooth muscle cells (VSMCs) and collagen. Elastin and the passive stress response of VSMCs are modelled as neo-Hookean materials, as standard practice \cite{Holzapfel_Gasser_Ogden_2000}. The active contraction of VSMCs is modelled according to the material model proposed by \cite{HumphreyCVS} where the functional shape of the stress response is selected according to experimental observations. Finally, the material model for collagen includes a distribution of levels of waviness of the collagen fibres, as experimentally observed in \cite{schrauwenetal} and experimentally validated in \cite{Hill2012}. 

The model captures three phases of the arterial vessel in relation to cerebral vasospasm: (1) healthy, pre-disease phase; (2) vasospasm at peak constriction; and (3) mechanical treatment. In the healthy phase all constituents are configured in their homeostatic state. In particular, their state of stretch equals a reference value called ``homeostatic stretch" which is the state of stretch they assume in healthy conditions. In a second stage, development of vasospasm is modelled via two mechanisms:
\begin{itemize}
    \item an incrase in vasoactive tone, driven by chemical signalling; this is modelled as an increase in the active stress response of VSMCs;
    \item remodelling of VSMCs which adjust their internal cytoskeleton and structure to the new, constricted vessel geometry; the remodelling mechanism is driven by the cells' effort to maintain homeostatic stretch (more details are given in Sections \ref{sec:methods-cm-mat}).
\end{itemize}

Finally, mechanical treatment is modelled as a gradual increase in internal pressure acting on the lumen (innermost layer) of the arterial wall. According to a strain-based damage model, we study when damage occurs across the thickness of the arterial wall and predict the magnitude of pressure necessary to damage the smooth muscle cells in all layers. 

The model is implemented in the finite element analysis program FEAP \cite{FEAPweb}, in an existing framework that had previously been developed to model aneurysm growth and remodelling \cite{eriksson2014, grytsan2015, grytsan2017}. The framework consisted of a rate-based constrained mixture approach and already included material models for elastin, ground matrix and collagen, as well as the option to model volumetric growth, isotropic or anisotropic, and fluid-solid-growth coupling. In order to integrate the vasospasm model, the following extensions were included in the framework:
\begin{itemize}
    \item introduction of material model for vascular smooth muscle cells, with both passive and active response;
    \item sophistication of material model for collagen to include fibre recruitment stretch distribution;
    \item remodelling of vascular smooth muscle cells;
    \item remodelling of collagen distribution;
    \item damage.
\end{itemize}

Details of the implementations of these extensions and their verification are given in \cite{pederzani2020}.

\subsection{Description of Finite Element Framework}
\label{sec:methods-framework}

\begin{figure}
    \centering
    \includegraphics[scale=0.7]{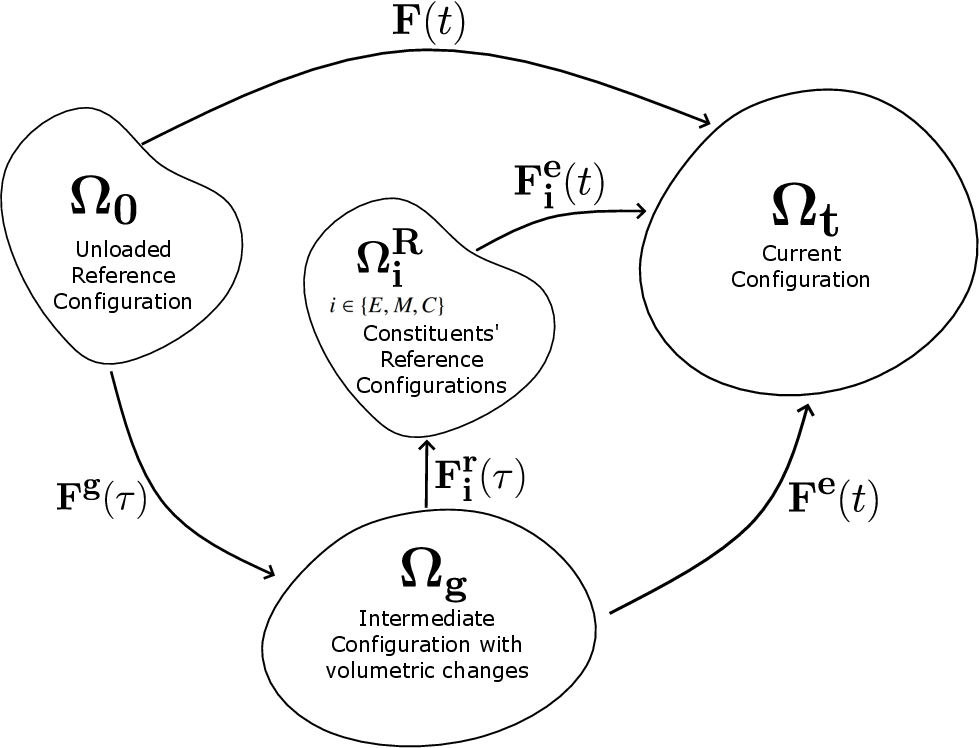}
    \caption{Split of deformation gradient $\mathbf{F}$ into growth ($\mathbf{F^g}$) and elastic ($\mathbf{F^e}$ ) parts, and corresponding states of constituent stretches in reference, intermediate and loaded configurations. *Collagen stretch is actually a distribution of values, but is shown here as a single value for simplicity.}
    \label{fig:def-gradient}
\end{figure}

The arterial wall is modelled as a single layer and the material is assumed incompressible and hyperelastic. The deformation gradient mapping the unloaded configuration to a current configuration is split into an elastic and a growth component, which allows modelling of volumetric growth (see Figure \ref{fig:def-gradient}). In other words:
\begin{equation}
    \mathbf{F} = \mathbf{F}^e \cdot \mathbf{F}^g,
\end{equation}
where $\mathbf{F}^e$ describes the elastic deformation and $\mathbf{F}^g$ the growth component (in our model $\mathbf{F}^g=\mathbb{I}_3$ identically). 

In order to numerically enforce incompressibility, the elastic deformation gradient is written as
\begin{equation}
    \mathbf{F}^e = (\det \mathbf{F}^e)^{\frac{1}{3}}\ \overline{\mathbf{F}},
\end{equation}
where $\det \overline{\mathbf{F}}=1$ by definition and $\det \mathbf{F}^e$ is numerically penalised for deviating from $1$ by postulating that the SEDF is of the form:
\begin{equation}
    \Psi = \overline{\Psi}(\mathbf{\overline{F}}) -p(\det \mathbf{F}^e -1),
\end{equation}
where $p$ is a Lagrange multiplier that can be identified as hydrostatic pressure. 

The strain energy density function $\overline{\Psi}(\mathbf{\overline{F}})$ is furthermore postulated to depend on functions of the elements of the right Cauchy-Green tensor $\mathbf{\overline{C}}=\mathbf{\overline{F}}^T\mathbf{\overline{F}}$, called invariants. These are given by Equations \eqref{invariants-iso}.
\begin{equation}
\begin{aligned}
    \label{invariants-iso}
    \overline{I}_1(\mathbf{\overline{C}}) =&\ tr(\mathbf{\overline{C}}), \\
    \overline{I}_2(\mathbf{\overline{C}}) =&\ \frac{tr(\mathbf{\overline{C}})^2 - tr(\mathbf{\overline{C}}^2)}{2}, \\
    \overline{I}_3(\mathbf{\overline{C}}) =&\ \det{\mathbf{\overline{C}}}, \\
    \overline{I}_4(\mathbf{\overline{C}},\mathbf{v}) =&\ \mathbf{v}\ \cdot\ \mathbf{\overline{C}} \mathbf{v}.
\end{aligned}
\end{equation}

The Definition of $\overline{I}_4(\mathbf{\overline{C}},\mathbf{v})$ in Equation \eqref{invariants-iso} can be adapted for any anisotropic material oriented along a single direction represented by unit vector $\mathbf{v}$. We will use $\overline{I}_4 = \overline{I}_4(\mathbf{\overline{C}},\mathbf{a_0})$ for collagen fibres aligned along unit vector $\mathbf{a_0}$, $\overline{I}_6 = \overline{I}_6(\mathbf{\overline{C}},\mathbf{g_0})$ for collagen fibres aligned along unit vector $\mathbf{g_0}$ and $\overline{I}_M = \overline{I}_M(\mathbf{\overline{C}},\mathbf{m_0})$ for vascular smooth muscle cells which are aligned along the circumferential direction $\mathbf{m_0}$.

The choices of material models of the microconstituents of the arterial wall, which are detailed in the next Sections, are such that the dependence on invariants $I_2, I_5, I_7$ and $I_8$ has been eliminated, while the dependence on $I_3$ has been substituted by the incompressibility constraint. Thus the constitutive equation determining the second Piola-Kirchhoff stress becomes
\begin{equation}
\label{simpconsteq}
\mathbf{S} = \: -p \mathbf{C}^{-1} + 2 \bigg[  \frac{\partial \overline{\Psi}}{\partial I_1} \mathbf{I} 
+ \frac{\partial \overline{\Psi}}{\partial I_4} \mathbf{A_0}  
+ \frac{\partial \overline{\Psi}}{\partial I_6} \mathbf{G_0}  
+ \frac{\partial \overline{\Psi}}{\partial I_M} \mathbf{M_0} \bigg],
\end{equation}
where $p$ is the hydrostatic pressure working as a penalty parameter that enforces incompressibility, and structure tensors $\mathbf{A_0}, \mathbf{G_0}$ and $\mathbf{M_0}$ are respectively given by $\mathbf{A_0} = \mathbf{a_0} \otimes \mathbf{a_0}$, $\mathbf{G_0} = \mathbf{g_0} \otimes \mathbf{g_0}$ and $\mathbf{M_0} = \mathbf{m_0} \otimes \mathbf{m_0}$.

\subsection{Homeostasis}
\label{sec:methods-cm-mat}

We adopt a constrained mixture approach where each microconstituent is modelled independently, has its own reference configuration, growth processes and remodelling mechanisms. In particular we assume independent reference configurations for each constituent ($\Omega_i^R$ in Fig. \ref{fig:def-gradient}). These reference configurations are defined according to the idea of \textit{homeostatic stretch}, namely a preferred state of stretch that each constituent aims to maintain in response to changes in the mechanical environment \cite{watton_mathematical_2004}. This allows the constituents to be in a different states of stretch in the ``healthy" artery depending on their properties and structural functions: this is well-documented for collagen where medial collagen is load bearing in health (thus has a stretch $\ge 1$, while adventitial collagen plays a protective role and does not bear load in healthy, so it appears wavy and its stretch is $<1$. The constituent-specific recruitment configuration $\Omega_i^R$ is then defined as the configuration at which the constituent is recruited to load bearing, in other words at which its stretch equals 1. With reference to Fig. 1 it will then hold that

\begin{equation}
\label{eq:recr-grad}
    \mathbf{F^e} = \mathbf{F_i^r} \cdot \mathbf{F_i^e},\ \forall i,
\end{equation}

where $\mathbf{F^e}$ contains the global deformation, $\mathbf{F_i^r}$ the constituent-specific recruitment deformation and $\mathbf{F_i^e}$ the constituent-specific deformation. For anisotropic components for which the problem reduces to one dimension, we will denote by $\lambda$ the global stretch resolved in the constituent direction, $\lambda_j^R$ the recruitment stretch (for constituent $j$), and $\lambda_j$ the constituent stretch. Equation \eqref{eq:recr-grad} will then become $\lambda = \lambda_j^R\ \lambda_j,\ \forall j$. 

For VSMCs a finite recruitment stretch $\lambda_M^R>1$ is chosen such that the cells' homeostatic stretch corresponds to a stress level lower than the maximum: this allows them to relax if necessary or to increase their active constriction if a change in the mechanical environment requires it (see Fig. \ref{fig:ch03-vsmc-active}).
For collagen a distribution of recruitment stretches is assumed: this allows the representation of the distribution of levels of stretch (or waviness) that the collagen fabric appears with \cite{schrauwenetal, Hill2012}. We assume a triangular distribution $\rho(\lambda_C^R)$ which is a good approximation of experimental observations (\cite{schrauwenetal}) and allows an analytical expression for the total collagen stress contribution, which avoids numerical integration and results in an efficient implementation with a simple remodelling scheme. The distribution is given by \cite{aparicio2016}:
\begin{equation}
\label{collagenpdftriangle}
\rho(\lambda_C^R) = \begin{cases}
0 & \text{\small if } \lambda_C^R < \lambda_{C}^{R,min}, \\
\frac{2 (\lambda_C^R - \lambda_{C}^{R,min})}{(\lambda_{C}^{R,max}-\lambda_{C}^{R,min}) (\lambda_{C}^{R,mod}-\lambda_{C}^{R,min})} & \text{\small if } \lambda_{C}^{R,min} \le \lambda_C^R < \lambda_{C}^{R,mod}, \\
\frac{2 (\lambda_{C}^{R,max} - \lambda_C^R)}{(\lambda_{C}^{R,max}-\lambda_{C}^{R,min}) (\lambda_{C}^{R,max}-\lambda_{C}^{R,mod})} & \text{\small if } \lambda_{C}^{R,mod} \le \lambda_C^R < \lambda_{C}^{R,max}, \\
0 & \text{\small if } \lambda_{C}^{R,max} \le \lambda_C^R.
\end{cases}
\end{equation}
In the above all stretches are to be understood as along the fibre direction, given by either $\mathbf{a_0}$ or $\mathbf{g_0}$ depending on the fibre family. We are assuming the same distribution for the two families of fibres.

\subsection{Material Models}

The stress response of the whole tissue is assumed to be the sum of the stress contributions from the individual microconstituents, i.e. 
\begin{equation}
\overline{\Psi} = \overline{\Psi}_{\mathrm{E}} + \overline{\Psi}_{\mathrm{C,a_0}} + \overline{\Psi}_{\mathrm{C,g_0}} + \overline{\Psi}_{M}^{pass} + \overline{\Psi}_{M}^{act},
\end{equation}
where the constituents are elastin ($\mathrm{E}$), two families of collagen fibres ($\mathrm{C}$) corresponding to two direction vectors along which they are aligned ($\mathbf{a_0},\ \mathbf{g_0}$), and vascular smooth muscle cells (M) with a passive (\textit{pass}) and an active (\textit{act}) response, the latter representing the cells' ability to relax and contract to regulate blood flow.

The constituent-specific strain energy density functions are selected as follows. 
Elastin is modelled as isotropic neo-Hookean material with SEDFs given by Equation \eqref{neohookean}
\begin{equation}
\label{neohookean}
        \overline{\Psi}_{\mathrm{E}} =\ \frac{k_\mathrm{E}}{2} (\overline{I}_1 - 3),
\end{equation}
where $k_\mathrm{E}$ a material parameter.

\subsubsection{Vascular Smooth Muscle Cells}
\begin{figure}
    \centering
    \includegraphics[scale=0.5]{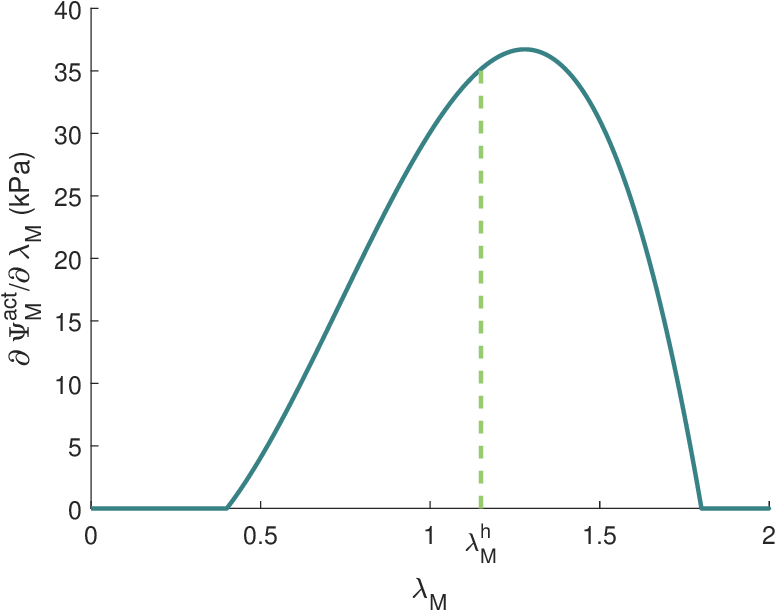}
    \caption{Active component of the stress response of vascular smooth muscle cells as a function of their stretch (solid curve). The dashed line corresponds to VSMC homeostatic stretch: $\lambda_{M}=\lambda_{M}^{h}=1.15$. }
    \label{fig:ch03-vsmc-active}
\end{figure}

Vascular smooth muscle cells are modelled as having both a passive and an active response. The passive response is modelled as neo-Hookean with SEDF given by
\begin{equation}
    \overline{\Psi}_{\mathrm{M}}^{pass} =\ \frac{k_\mathrm{M}^{pass}}{2} (\overline{I}_1 - 3),
\end{equation}
where $k_\mathrm{M}^{pass}$ is a material parameter.
The active stress response is inferred from experimental measurements reported in \cite{Rachev_Hayashi_1999} and taken from Baek et al. \cite{Baek2007}. It is non-zero only in the circumferential direction and has a functional form such that it is non-zero only in interval $(\lambda_{M}^{min}, \lambda_{M}^{max})$ and the position of the maximum within that interval is controlled by $\lambda_{M}^{mean}$ (\cite{Ushiwata1990,Rothermel2020}).  It is given by Equation \eqref{eq-vsmc-act}:
\begin{equation}
\label{eq-vsmc-act}
    \frac{\partial\Psi_{M}^{act}}{\partial \lambda_M} =\ k_A\ c_v\ k_{M}^{act}\ \lambda_{M}\ \left[ 1 - \left( \frac{\lambda_{M}^{mean} - \lambda_{M}}{\lambda_{M}^{mean}-\lambda_{M}^{min}} \right)^2 \right],
\end{equation}
where $k_A$ is a parameter controlling the increase of the above ratio in vasospasm ($k_A=1$ at $t=0$), $c_v$ is the physiological ratio of vasoconstrictors to vasodilators concentrations, $k_{M}^{act}$ is a material parameter, parameters $\lambda_M^{min}, \lambda_M^{mean}$ and $\lambda_M^{max}$ control the shape of the active stress response \cite{Baek2007}, and  $\lambda_M$ is the cell stretch in the circumferential direction $\left(\lambda_M = \sqrt{\overline{I_M}/I_M^R}\right)$, with $I_M^R = (\lambda_M^R)^2$ (recall Eq. \eqref{eq:recr-grad}). The plot of $\frac{\partial\Psi_{M}^{act}}{\partial \lambda_M}$ is shown in Fig. \ref{fig:ch03-vsmc-active} for the parameter values used in this model and reported in Table \ref{cylcvspar}.

\subsubsection{Collagen}
For collagen we adopt a constitutive model that includes the stretch distribution defined in Eq. \eqref{collagenpdftriangle}. This requires the definition of a SEDF for an individual fibre, which is taken from Chen \cite{chenIntracranialAneurysmDisease2014} and Aparìcio et al. \cite{aparicio2016} as a linear stress response in the first Piola-Kirchhoff stress: 
\begin{equation}
    \tilde{\Psi}_C (\lambda_C) = \frac{\mu_C}{2} \left(\lambda_C - 1\right)^2,
\end{equation}
where $\mu_C$ is a stiffness-like material constant. The collagen stretch $\lambda_C$ is given by $\lambda_C = \sqrt{\overline{I_4}(\mathbf{\overline{C}},\mathbf{a_0}) / (\lambda_C^R)^2) }$, with $\lambda_C^R$ varying according to $\rho(\lambda_C^R)$.

The total stress contribution from the collagen material is then obtained by integrating over the distribution of recruitment stretches:
\begin{equation}
    \label{eq:collagen_integral}
    \Psi_C(\lambda) = \int_1^{\lambda} 
    \tilde{\Psi}_C\left(\frac{\lambda}{\lambda_C^R}\right)\  \rho(\lambda_C^R)\ 
    d\lambda_C^R.
\end{equation}
The choice of a triangular distribution results in the first Piola-Kirchhoff stress admitting an analytical piecewise continuous form:
\begin{equation}
    \frac{\partial \Psi_C}{\partial \lambda}(\lambda) = \mu_C\ C_1 \left[ (\lambda + C_2) \ln(\frac{\lambda}{C_3}) + C_4\ \lambda + C_5  \right],
\end{equation}
where $C_i = C_i(\lambda_C^{R,min},\lambda_C^{R,mod},\lambda_C^{R,max})$ are constants depending on $\lambda_C^{R,min},\lambda_C^{R,mod}$ and $\lambda_C^{R,max}$.

The SEDFs for the two families of collagen fibres are identical apart from the invariant from which the fibre stretch is computed. Indeed for the family oriented along $\mathbf{a_0}$, $\lambda = \sqrt{\overline{I_4}(\mathbf{\overline{C}},\mathbf{a_0})}$ and analogously for the family oriented along $\mathbf{g_0}$, $\lambda = \sqrt{\overline{I_6}(\mathbf{\overline{C}},\mathbf{g_0})}$, where $\lambda$ is the tissue stretch resolved in the direction of the fibres (recall $\lambda = \lambda_C \cdot \lambda_C^R$ from Eq. \eqref{eq:recr-grad} ).

\subsection{Growth and Remodelling}
\label{sec:methods-gandr}

In the model proposed here growth processes are assumed to be negligible, while remodelling of constituents plays a key role. In particular we assume that a second phase of vasospasm is dominated by VSMC remodelling to restore a homeostatic level of stretch. This can occur via cytoskeletal reconfiguration and/or changes in protein-mediated attachments to neighboring cells and/or the surrounding matrix. The modelling framework also accommodates remodelling of collagen, although this process is assumed to be negligible in this application to vasospasm. 

Both remodelling processes are characterised by the same dynamic, namely the constituent aiming to maintain its current stretch close to its homeostatic value. In the case of VSMCs, this is given by Equation \eqref{eq-vsmc-remodel}:
\begin{equation}
\label{eq-vsmc-remodel}
    \frac{\partial \lambda_M^R}{\partial t} = \alpha_M \frac{ \lambda_M - \lambda_M^{h} }{\lambda_M^{h}},
\end{equation}
where $\lambda_M^{h}$ is the target homeostatic value.

The case of collagen must be adapted because of the inclusion of a distribution of stretches. A fibre stretch distribution is defined that corresponds to physiological homeostatic conditions:
\begin{equation}
\label{cud_triplet_hom}
        \left( \lambda_C^{min,h}, \lambda_C^{mod,h}, \lambda_C^{max,h} \right).
\end{equation}

Collagen can remodel towards this target distribution according to the following equations, which are linear in the deviation from the corresponding target stretches:
\begin{equation}
\begin{aligned}
\label{eq:remodel-coll-dist}
    \frac{\partial \lambda_{C}^{R,min}}{\partial t} =&\ \alpha_C 
    \frac{\lambda_C^{max} - \lambda_C^{max,h}}{\lambda_C^{max,h}}, \\   
    \frac{\partial \lambda_{C}^{R,mod}}{\partial t} =&\ \alpha _C
    \frac{\lambda_C^{mod} - \lambda_C^{mod,h}}{\lambda_C^{mod,h}}, \\ 
    \frac{\partial \lambda_{C}^{R,max}}{\partial t} =&\ \alpha_C 
    \frac{\lambda_C^{min} - \lambda_C^{min,h}}{\lambda_C^{min,h}},
\end{aligned}
\end{equation}
where $\lambda_C^{min}, \lambda_C^{mod}$ and $\lambda_C^{max}$ are the current collagen stretches while $\lambda_{C}^{R,min}, \lambda_{C}^{R,mod}$ and $\lambda_{C}^{R,max}$ the current recruitment stretches. Thus, if for example $\lambda_C^{max}>\lambda_C^{max,h}$, then the partial derivative of $\lambda_{C}^{R,min}$ is positive and the system will evolve to increase $\lambda_{C}^{R,min}$, resulting in the desired decrease in $\lambda_C^{max}$. Similarly for $\lambda_{C}^{R,mod}$ and $\lambda_{C}^{R,max}$.

\subsection{Model of Cerebral Vasospasm} 
\label{sec:methods-fecvs}

The development of vasospasm is modelled according to the following:
\begin{itemize}
    \item the increase in vasoconstrictors and scavengers of vasodilators, result of a signalling cascade initiated by the extravascular blood clot, causes an increase in the active tone of VSMCs (which, due to the higher stress, results in a decrease in luminal diameter);
    \item following this chemically-driven constriction, the smooth muscle cells remodel to return their level of stretch to its homeostatic value;
    \item elastin and collagen do not remodel;
    \item no growth or atrophy occurs.
\end{itemize}

The increase in the active stress response of VSMCs is achieved by pre-multiplication of their stress function by a parameter $k_A$ (see Eq. \eqref{eq-vsmc-act}), which is set equal to $1$ in healthy conditions and increases in vasospasm. The constriction is modelled such that it is maximal in the middle of the geometry and negligible at the ends. This is achieved by increasing $k_A$ towards a target value that is a Gaussian function of the axial coordinate of each element \eqref{eq:kA-gaussian}:
\begin{equation}
    \label{eq:kA-gaussian}
    k_A^{target} = 1 + k_A^{max} \cdot \exp\left({-\frac{1}{2}\:\left(\frac{z - z_{mid}}{\sigma}\right)^2}\right),
\end{equation}
where $z_{mid}$ is the axial material coordinate of the middle of the artery. Values of the other parameters are reported in Table \ref{tb:par-cvs}.

The increase of $k_A$ towards the target value is prescribed by a simple ODE \eqref{eq:kA-increase}:
\begin{equation}
    \label{eq:kA-increase}
    \frac{\partial k_A}{\partial t} = k_A^{target} - k_A.
\end{equation}

In parallel to the increase in active stress, the evolution equations for remodelling of VSMCs are run, which describe the cells' remodelling to restore the homeostatic value of stretch:
\begin{equation}
    \frac{\partial \lambda_M^R}{\partial t} = \alpha_M \frac{ \lambda_M - \lambda_M^{h} }{\lambda_M^{h}}.
\end{equation}

\begin{table}[h!]
	\centering
	\begin{tabular}{| c | c |}
		\hline
		Parameter & Value \\
		\hline
		$k_A^{max}$ & $17$ \\
        $z_{mid}$   & $0.625$mm \\
        $\sigma$    & $0.002$ \\
        $\alpha_M$  & $2$ \\
		\hline
	\end{tabular}
\caption{ \label{tb:par-cvs} Table of relevant parameters for the simulation of cerebral vasospasm. }
\end{table}

\subsection{Damage}
\label{sec:methods-damage}

Following \cite{Li2012} a scalar damage variable $d_{0,i}\in[0,1]$, with $i\in\{ E,C,M\}$ is introduced for each constituent, which represents the extent of damage. The case $d_{0,i}=0$ corresponds to no damage, while $d_{0,1}=1$ corresponds to complete failure and the cessation of load bearing. Indeed, the strain energy density functions of the individual constituents are pre-multiplied by the quantity $(1-d_{0,i})$ and thus damage to a constituent corresponds to a decrease in its contribution to load-bearing:
\begin{equation}
    \Psi_i \longrightarrow (1-d_{0,i}) \Psi_i,
\end{equation}
for $i\in\{ E,C,M^{pass},M^{act}\}$.

In other words, the total SEDF of the tissue is now given by
\begin{equation}
    \Psi  =  (1-d_{0,E}) \Psi_E + (1-d_{0,C}) \Psi_C +  (1-d_{0,M}) (\Psi_M^{pass} +  \Psi_M^{act}). 
\end{equation}

For our modelling purposes, it is only necessary to have a damage criterion for VSMCs, but it would be straight-forward to extend the framework to accommodate criteria for all the damage variables.

A strain-based damage criterion is used: it is assumed that there is a level of stretch $\lambda_{M,d}^{min}$ (with $I_{M,d}^{min}=(\lambda_{M,d}^{min})^2$), called \textit{minimum damage threshold}, which if exceeded, initiates damage. If the cell stretch remains above this threshold, damage is increased at each numerical step by a factor proportional to the deviation of the cell stretch from the minimum damage threshold. If the cell continued to be in overextension, damage would increase at every time step. It is ensured that $d_{0,M}$ never exceeds one.

In other words:
\begin{equation}
\label{eq:fe-dmg-crit}
    d_{0,M} = \min(\: d_{0,M} + \alpha_d (I_{M} - I_{M,d}^{min}),\: 1)\ \mathbf{1}_{I_{M} \ge I_{M,d}^{min}}, 
\end{equation}

where $I_{M}=(\lambda_M)^2$ and $\mathbf{1}$ is the indicator function (i.e. equal to $1$ when the subscript condition is satisfied, $0$ otherwise).

The choice of parameters $I_{M,d}^{min}$ and $\alpha_d$ is driven by the experimental results reported in \cite{Fischell1990} which suggest an $80-90\%$ stretch as a threshold for VSMC injury. In our simulation the level of cell stretch at which damage occurs will depend on the radial coordinate and on the level of stenosis, so the parametrisation aimed at obtaining a range of values approximately centered around $1.85$.

\subsection{Mechanical Treatment}
\label{sec:methods-mech-treat}

Mechanical treatment is modelled as an increase in internal pressure. The implementation makes use of an inbuilt function of the FEAP software which allows the gradual increase of the internal pressurisation up to a prescribed fraction of the boundary condition value. We set this gradual increase to achieve an additional pressure of 12$kPa$ which is sufficient to achieve damage in our simulations. This additional pressure is applied equally to all elements, so it does not differentiate between the location of the spasm and the areas of physiological geometry. During this phase, we assume no remodelling because the time of mechanical intervention, which is in the order of seconds or minutes, is significantly smaller than the time scale of the remodelling processes.

In order to test whether this additional pressure would be sufficient to bring VSMCs to functional failure, which is the success criterion for an interventional device, the damage model described in Section \ref{sec:methods-damage} is used.

The model is divided into two parts: the first part captures the transformation of the geometry and its microstructure through four key configurations (unloaded, loaded, health and vasospasm), while the second part simulates mechanical treatment of the vasospastic vessel.

\subsection{Geometry and Boundary Conditions}
\label{sec:methods-BCs}

The geometry is modelled as a quarter cylinder with 48 elements in the axial direction, 8 in the circumferential and 8 in the radial. The elements are 8-node bricks of type Q1P0.

The boundary conditions are applied to replicate the \textit{in vivo} conditions of cerebral arteries: they consist of an internal pressurisation to a $p=16kPa$, corresponding to physiological systolic blood pressure, and an axial pre-stretch of $\lambda_z$ (see Table \ref{cylcvspar}). 

The boundary conditions in Eq. \eqref{eq-disp-bc} are applied to the cylinder, where the first three conditions fix the boundary surfaces $x=0$, $y=0$ and $z=0$, so that the geometry can be thought of as the symmetric quarter of a cylinder, while the latter condition enforces the axial pre-stretch of $\lambda_z=1.2$ (note that units are in meters, so $0.015m = 1.5cm$).
\begin{equation}
\begin{aligned}
\label{eq-disp-bc}
     u_x(x=0)=&\ 0, \\
     u_y(y=0)=&\ 0, \\
     u_z(z=0)=&\ 0, \\
    u_z(z=0.0125)=&\ 0.015.
\end{aligned}
\end{equation}

Material parameters and other parameters relevant to the constitutive models of the constituents are reported in Table \ref{cylcvspar}.
\begin{table}[h!]
	\centering
	\begin{tabular}{| c | c |}
		\hline
		Parameter & Value \\
		\hline
		H & 0.29mm \\
		R & 1.45mm \\
		H/R & $1/5$ \\
		p & $16kPa$ \\
		$\lambda_z$ & $1.2$ \\
		$k_E$ & $93 kPa$ \\
		$k_C$ & $5800 kPa$ \\
		$k_{M}^{pass}$ & $45.1 kPa$ \\
		$k_{M}^{act}$ & $11 kPa$ \\
        $\phi_{\mathbf{a_0}}$ & $\frac{\pi}{4}$ \\
        $\phi_{\mathbf{g_0}}$ & $-\frac{\pi}{4}$ \\
        $\lambda_M^{h}$ & $1.15$ \\
        $\lambda_C^{h,min}$ & $0.85$ \\
        $\lambda_C^{h,mod}$ & $0.95$ \\
        $\lambda_C^{h,max}$ & $1.05$ \\
        $\alpha_M$ & $2$ \\
		\hline
	\end{tabular}
\caption[Model parameters of finite element model of cerebral vasospasm]{ \label{cylcvspar} Table of relevant model parameters for finite element model of cerebral vasospasm. }
\end{table}

\subsection{Parameter Studies}
\label{sec:methods-parstudy}

Finally we include two parameter studies: in the first we compare the model predictions for arteries of different original diameters, while in the second we evaluate the effect of different values of arterial thickness on the pressure predictions. 

To compare arteries of different original diameters we consider values of $2$, $2.9$ and $4$mm based on reported uses of stent-assisted angioplasty in cerebral vessels of about these values. Clinical observations suggest $\tilde 3$mm as a threshold for the effectiveness of currently available stent-retrievers, thus we wish to test whether our results predict a similar threshold.

For the wall thickness comparison, we are interested in the parameter $H/R$, namely the unloaded thickness-to-radius ratio. We found two studies that report \textit{ex vivo} measurements of arterial thickness and radius, both showing a varied distribution. In \cite{Monson_Barbaro_Manley_2008} 12 specimens of cerebral arteries were measured and axially tested, all with $<1mm$ diameter and presumably healthy, with patient age range 16-62 years (mean 37): the resulting range for $H/R$ is 0.25-0.35. In \cite{Harteveld_2018} 10 specimens from the Circle of Willis were studied, with and without cardiovascular disease, with patient age range 66-84 years (mean 76): the resulting range for $H/R$ is 0.30-0.50. This suggests significant interpersonal variability for this parameter, which may be affected by patient age and presence/severity of cardiovascular disease. In the main results we use a value of 0.35 as an average. In the parameter study we vary this parameter $H/R \in \{ 0.20, 0.25, 0.30, 0.35 \}$ and compare the predictions of pressure requirements for the mechanical resolution of vasospasm.

\section{Results}
\label{sec:results}

After an initial stage that achieves the configuration corresponding to an artery before vasospasm (via 1. application of boundary conditions, and 2. remodelling to mechanical homeostasis), the model captures the evolution of the arterial geometry and the configurations of its constituents through three relevant configurations: 
\begin{enumerate}
    \item \textbf{pre-vasospasm}: artery is pre-stretched and pressurised, all constistuents are in mechanical homeostasis;
    \item \textbf{vasospasm}: after increase of active contraction and (complete) remodelling to restore homeostatic stretch;
    \item \textbf{treatment}: after application of additional internal pressure with damage to VSMCs.
\end{enumerate}
Section \ref{subsec:res-healthy} reports the results for the transformation of the system from the unloaded configuration to the pre-vasospasm homeostatic state; Section \ref{subsec:fecvs-res-vaso} reports the results regarding the development of vasospasm; and finally Section \ref{subsec:fecvs-res-treat} is concerned with the results relating to treatment.

\subsection{Artery before Vasospasm}
\label{subsec:res-healthy}

The transformation of the meshed geometry into a state corresponding to the artery before vasospasm in \textit{in vivo} conditions consists of two parts:
\begin{enumerate}
    \item the reference (unloaded) configuration is first pressurised with an internal pressure of $16 kPa$ (physiological systolic blood pressure) and axially pre-stretched; this achieves the Loaded configuration;
    \item the remodelling routines are run so that VSMCs and collagen achieve their homeostatic configurations thus realising a state of physiological homeostasis and capturing the mechanically homeostatic configuration of the arterial wall.
\end{enumerate}
Figures \ref{fig:timelapse-lamm} and \ref{fig:timelapse-lammr} show the geometry, VSMC stretch (Fig. \ref{fig:timelapse-lamm}) and VSMC recruitment stretch (Fig. \ref{fig:timelapse-lammr}) for the three configurations corresponding to unloaded, pre-vasospasm and vasospasm configurations. Note that the transition from unloaded to pre-vasospasm configuration is composed of two steps, the application of boundary conditions and the remodelling towards mechanical homeostasis, but these are combined for clarity. After application of the boundary conditions, the recruitment stretch remodels so that VSMC stretch equals its homeostatic value and thus a uniform transmural strain field is realised in the homeostatic pre-vasospasm configuration (second row).

\subsection{Vasospasm}
\label{subsec:fecvs-res-vaso}

\begin{figure}[h]
    \centering
    \includegraphics[scale=0.89]{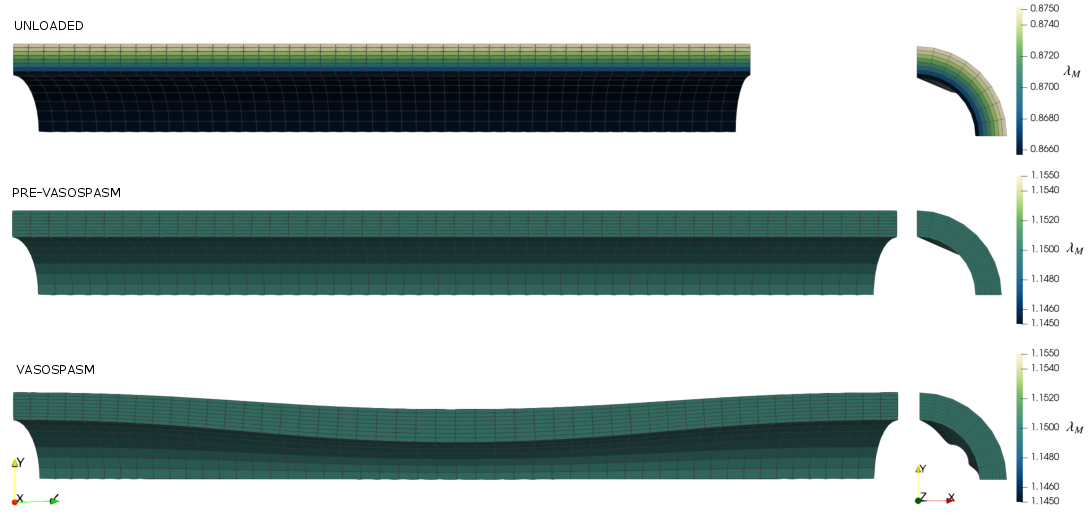}
    \caption{Axial and radial views of VSMC stretch $\lambda_M$ distribution at three representative stages of the simulation: unloaded geometry, pre-vasospasm and vasospasm. Following loading from the stress-free configuration, VSMCs remodel until their stretch equals the homeostatic value uniformly across the geometry. Since remodelling to maintain homeostatic stretch continues during vasospasm, the uniform strain field remains in this phase as well.}
    \label{fig:timelapse-lamm}
\end{figure}

\begin{figure}[h]
    \centering
    \includegraphics[scale=0.89]{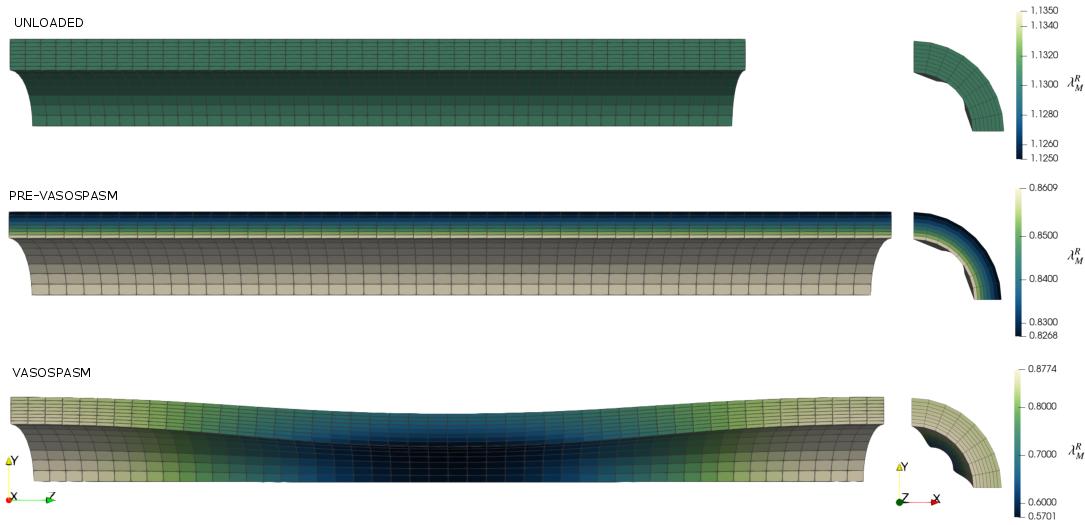}
    \caption{Axial and radial views of VSMC recruitment stretch $\lambda_M^R$ distribution at three representative stages of the simulation: unloaded geometry, pre-vasospasm state and vasospasm. Following loading, VSMCs remodel until their stretch equals the homeostatic value uniformly across the geometry, resulting in a radial gradient of the recruitment stretch. Following vasospasm, the new geometry requires a recruitment stretch distribution that varies both radially and axially.}
    \label{fig:timelapse-lammr}
\end{figure}

From the pre-vasospasm configuration the arterial geometry is transformed into a constricted, vasospastic state with a peak stenosis of $~36\%$ in the middle of the vessel. This is achieved by increasing the active tone of vascular smooth muscle cells and by continuing the cells' remodelling to maintain homeostatic stretch. The combined effect of the increased active stress and recruitment stretch remodelling causes a local stenosis of the vessel until peak constriction is achieved in the Vasospasm configuration (Figures \ref{fig:timelapse-lamm} and \ref{fig:timelapse-lammr}, bottom row), at which point VSMC stretch has returned to its homeostatic value across the wall thickness.

Figure \ref{fig:timelapse-lammr} highlights the wide range of VSMC recruitment stretch values needed to accommodate homeostatic cell stretch through the change in geometry. While before vasospasm only a gradient in the radial direction was necessary, in vasospasm the recruitment stretch has attained a distribution of values with gradients across both the axial and radial directions. The addition of a gradient in the axial direction is indeed necessary to maintain the cell stretch in homeostasis when the geometry is no longer uniform in the axial direction. 

\begin{figure}
    \centering
    \includegraphics[scale=0.6]{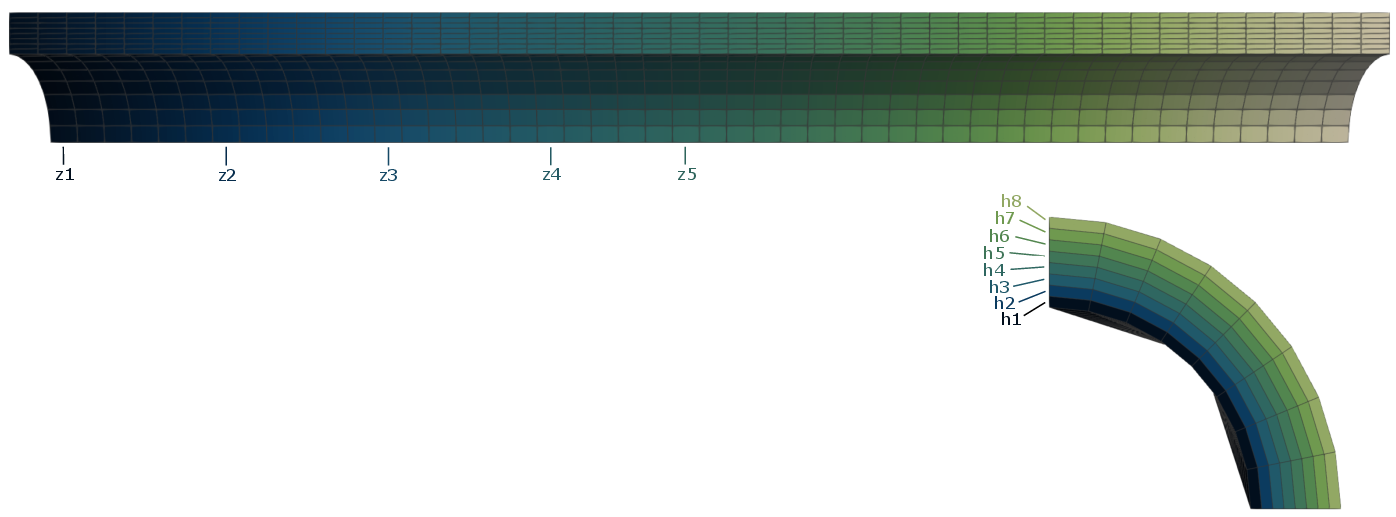}
    \caption{Identification of representative locations on vessel geometry for interpretation of results.}
    \label{fig:elemid-innout}
\end{figure}

\begin{figure}
    \centering
    \includegraphics[scale=0.8]{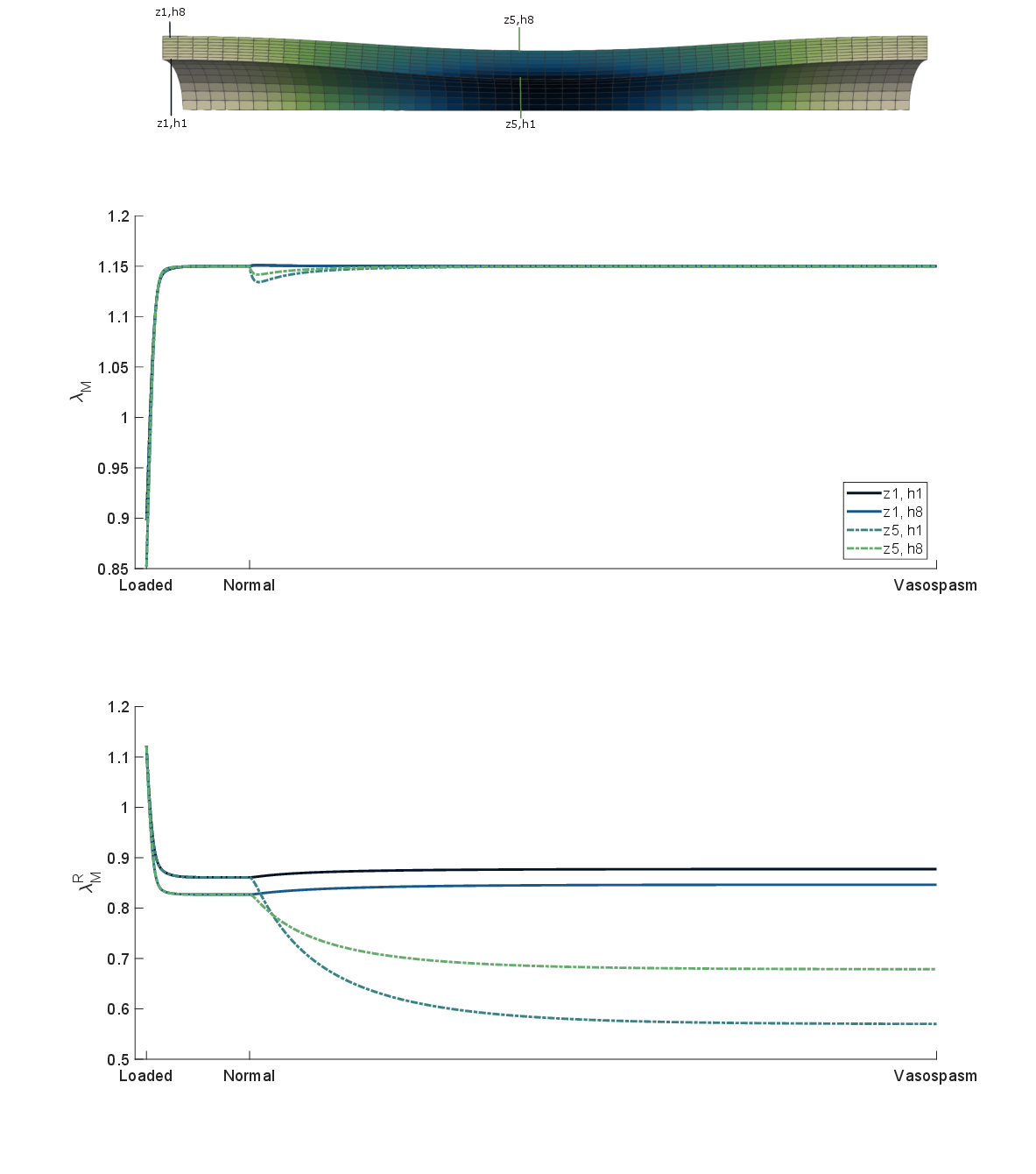}
    \caption{Evolution of VSMC cell stretch (top) and recruitment stretch (bottom) in four representative locations of the artery during the development of vasospasm. Above are shown the four representative locations for which the values are shown. At vasospasm inception, the increase in active contraction causes a deviation of cell stretch from its homeostatic value. Over time remodelling of recruitment stretch returns cell stretch to homeostasis in all elements, though this requires a wide range of recruitment stretch values to maintain uniformity through the deformed geometry. }
    \label{fig:fecvs-stretches-plot}
\end{figure}

The quantitative evolution of $\lambda_M$ and $\lambda_M^R$ is displayed in Figure \ref{fig:fecvs-stretches-plot} on a continuous-in-time scale. The values of cell and recruitment stretches are reported for four representative elements in the geometry. Throughout the rest of the manuscript we will use the labelling system shown in Figure \ref{fig:elemid-innout} to report results in various relevant locations of the geometry. In Figure \ref{fig:fecvs-stretches-plot} the four elements are chosen to compare elements in the ``unaffected" part of the geometry ($z1$) versus the area of peak constriction ($z5$), and in the innermost layer ($h1$) versus the outermost ($h8$). 

During development of vasospasm, VSMC contraction and remodelling occur. Looking at cell stretch $\lambda_M$, it is visible how at the inception of vasospasm  the increase in vasoactive tone is the dominating factor which causes an initial deviation of VSMC stretch from the homeostatic value. As remodelling ``catches up" with the constricting geometry, cell stretches across all elements are gradually returned to the homeostatic value ($\lambda_M^h=1.15$). Looking at the recruitment stretches $\lambda_M^R$, it is visible how they adapt to the changing geometry and how wide a range of values is required to maintain homeostatic cell stretch through the deformed geometry. It is interesting to note how at $z1$ the values of recruitment stretch are almost unchanged: they increase slightly but their difference remains similar. By contrast in the area of peak vasospasm ($z5$) the recruitment stretches deviate significantly from their values in healthy homeostasis and the through-thickness difference is also wider. It is evident that the dependence on the axial direction, therefore on the severity of vasospasm, is stronger than the dependence on the location in the radial direction.

\subsection{Treatment}
\label{subsec:fecvs-res-treat}

\begin{figure}
    \centering
    \includegraphics[scale=0.82]{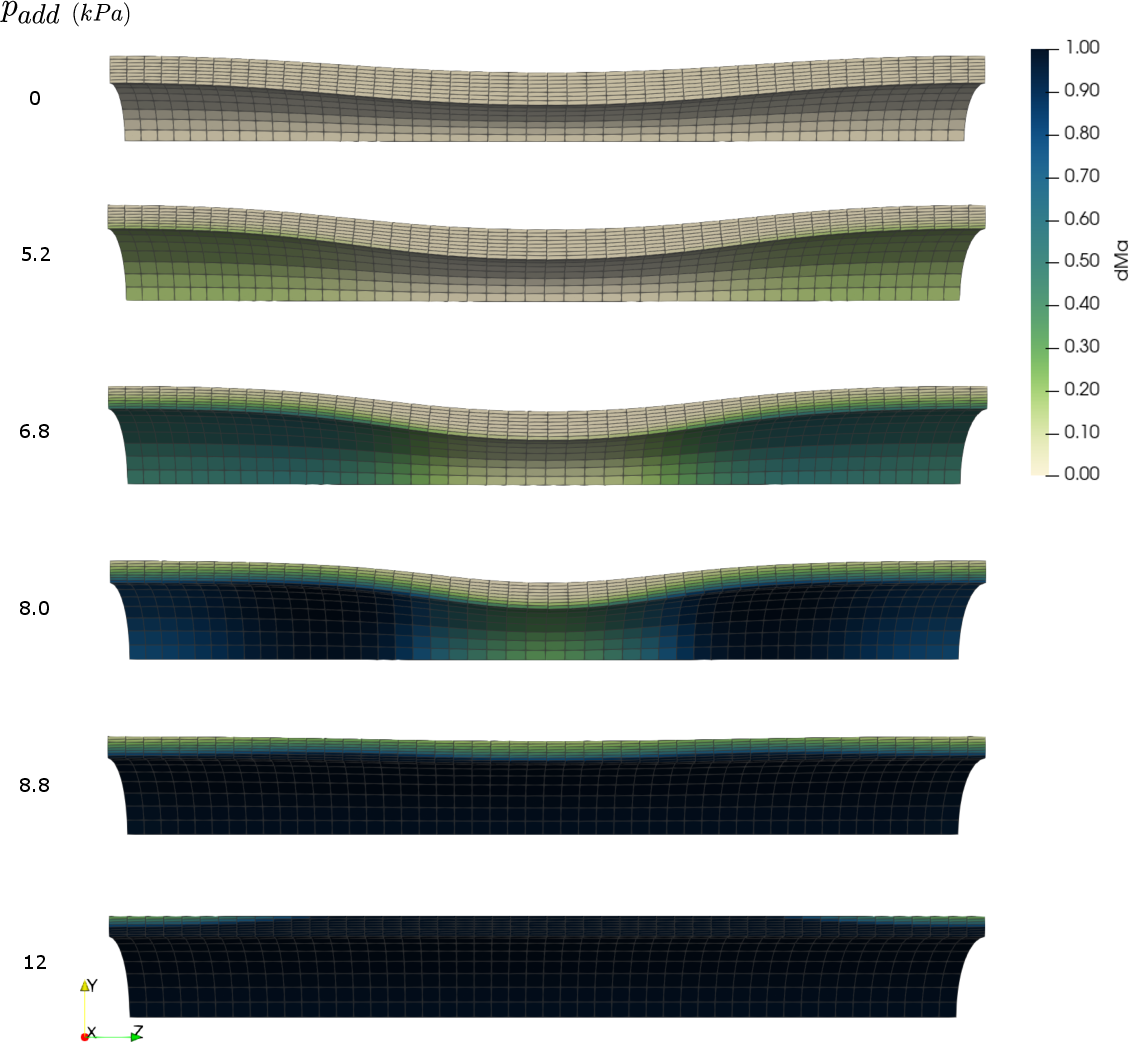}
    \caption{Propagation of damage through geometry during simulation of vasospasm treatment at selected qualitatively relevant moments. At $p_{add}=0$ kPa the configuration is at beginning of treatment. At $p_{add}=5.2$ kPa initial damage is visible at the ends of the geometry, then it slowly propagates towards the middle ($p_{add}=6.8$ kPa). At $p_{add}=8$ kPa the highest damage occurs in the transition areas between the ``unaffected" (axial extremities) and the constricted tissue. At $p_{add}=8.8$ kPa the innermost layer is completely damaged and the fastest propagation begins to move from the transition areas to the middle of the artery. At the end of treatment ($p_{add}=12$ kPa) all layers are damaged in the central part of the artery while at the axial extremities the outer layers do not reach complete damage, consistently with the distribution of the VSMC recruitment stretch.}
    \label{fig:dma-snaps}
\end{figure}

The second part of the simulation is concerned with modelling mechanical treatment of vasospasm. This is modelled via the application of an additional internal pressure to the innermost layer, which is linearly increased from $0$ to $12kPa$. Figure \ref{fig:dma-snaps} shows snapshots of the geometry as damage propagates due to the pressure increase.

\begin{figure}
    \centering
    \includegraphics[scale=0.64]{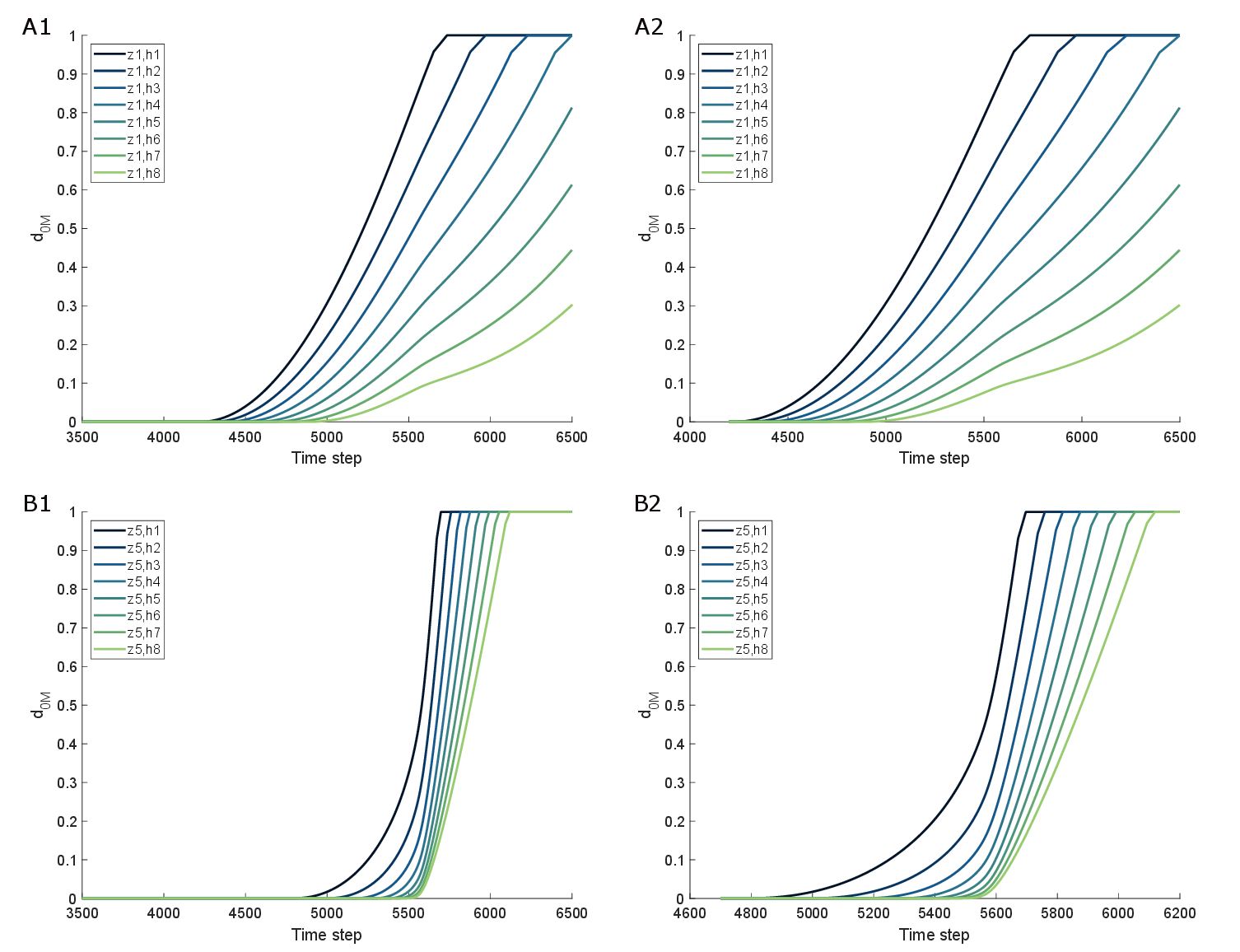}
    \caption{Evolution of damage variables during treatment of vasospasm at the ``unaffected" end (z1, top row) and in the area of peak constriction (z5, bottom row). Plots in the left column comprise the whole time interval of treatment, while the right column excludes times where the damage variable is 0 or 1 for all elements. }
    \label{fig:cyldam_allelem}
\end{figure}

The evolution of the damage variables $d_{0,M}$ is shown in Figure \ref{fig:cyldam_allelem} where the top row shows the propagation at the ``normal" end of the geometry (at $z_1$ with reference to Fig. \ref{fig:elemid-innout}) and the bottom row in the area of peak constriction (at $z5$). The second column restricts the visualization to time steps at which the damage variables are not all $0$ or $1$. Damage initiates first in the innermost layer propagates through the wall thickness until the outermost layer is the last to be damaged. As expected, damage initiates in each layer when the damage threshold is reached and a previous layer doesn't need to be completely damaged before partial damage can begin in the next one. It is also interesting to note different behaviour at the two longitudinal locations considered: in the ``unaffected" end of the artery damage begins sooner but propagates slower and at the end of the treatment simulation four layers have not reached complete damage. Conversely, in the area of peak constriction, damage begins later but propagates significantly faster through the wall thickness and all layers are completely damaged before the simulation is completed. This is consistent with the distribution of VSMC recruitment stretches shown in Fig. \ref{fig:timelapse-lammr}, as the lower values of recruitment stretch in the middle of the artery result in higher constituent stretches for the same global deformation, therefore VSMCs in this area experience higher constituent stretches and thus damage propagates faster compared to the longitudinal extremities of the geometry where the values of recruitment stretch remained close to the pre-vasospasm distribution.

It is of interest to focus on the area of peak constriction to study the propagation of damage and its relationship with the corresponding amount of additional internal pressure applied. Points of interest are either times at which damage is initiated, i.e. $\min_t d_{0,M}(t)>0$, or times at which complete functional failure is achieved, i.e. $\min_t d_{0,M}(t)=1$.  Table \ref{tbl:fecvs-dmg_pres} reports these values for each through-thickness element (numbered as shown in Fig. \ref{fig:elemid-innout}), together with the corresponding levels of additional pressure and the corresponding level of VSMC stretch. These values are extracted from the data set plotted in Fig. \ref{fig:cyldam_allelem}. Please note that ``time" here is used in reference to the numerical step, but does not have true temporal value.

\begin{table}
    \centering
    \begin{tabular}{c|c|c|c|c|c|c}
     & $\mathbf{t_0}$ & $\mathbf{p_{add}(t_0)} $ &  & $\mathbf{t_1}$ & $\mathbf{p_{add}(t_1)} $ & \\
    \textbf{h-el} & $\mathbf{\min_t (d_{0,M}(t)\!>\!0)}$ & $ (kPa)$ & $\mathbf{\lambda_{M_{|t_0}}}$ & $\mathbf{\min_t (d_{0,M}(t)\!=\!1)}$ & $ (kPa)$ & $\mathbf{\lambda_{M_{|t_1}}}$ \\
    \hline
     h1  & 4767 & 5.07 & 1.2980 & 5698 & 8.79  & 2.1576 \\
     h2  & 4973 & 5.89 & 1.2987 & 5760 & 9.04  & 2.0643 \\
     h3  & 5143 & 6.57 & 1.2992 & 5818 & 9.27  & 1.9782 \\
     h4  & 5270 & 7.08 & 1.2995 & 5875 & 9.50  & 1.9034 \\
     h5  & 5360 & 7.44 & 1.3000 & 5933 & 9.73  & 1.8394 \\
     h6  & 5422 & 7.69 & 1.3005 & 5992 & 9.97  & 1.7842 \\
     h7  & 5465 & 7.86 & 1.3008 & 6054 & 10.22 & 1.7358 \\
     h8  & 5497 & 7.99 & 1.3018 & 6119 & 10.48 & 1.6922
    \end{tabular}
    \caption{Evolution of damage variables across the eight through-thickness elements in the area of peak constriction. For each element are reported: the time at which damage initiates ($\min_t d_{0,M}(t)>0$), the corresponding additional pressure and cell stretch; the time at which damage is complete ($\min_t d_{0,M}(t)=1$), the corresponding additional pressure and cell stretch.}
    \label{tbl:fecvs-dmg_pres}
\end{table}

These results highlight the relationship between the additional internal pressure and damage propagation. For an artery of physiological diameter $2.9mm$ at about $36\%$ peak stenosis the prediction is that:
\begin{itemize}
    \item $\sim5 kPa$ of additional pressure are necessary to initiate damage in the innermost layer,
    \item $\sim8 kPa$ are necessary to initiate damage in all layers,
    \item $\sim9 kPa$ to bring VSMCs in the innermost layer to $d_{0,M}=1$, and
    \item $\sim10.5 kPa$ to bring VSMCs to functional failure across all layers.
\end{itemize}

\subsection{Parameter Study: Original Artery Diameter}

\begin{figure}
    \centering
    \includegraphics[scale=0.6]{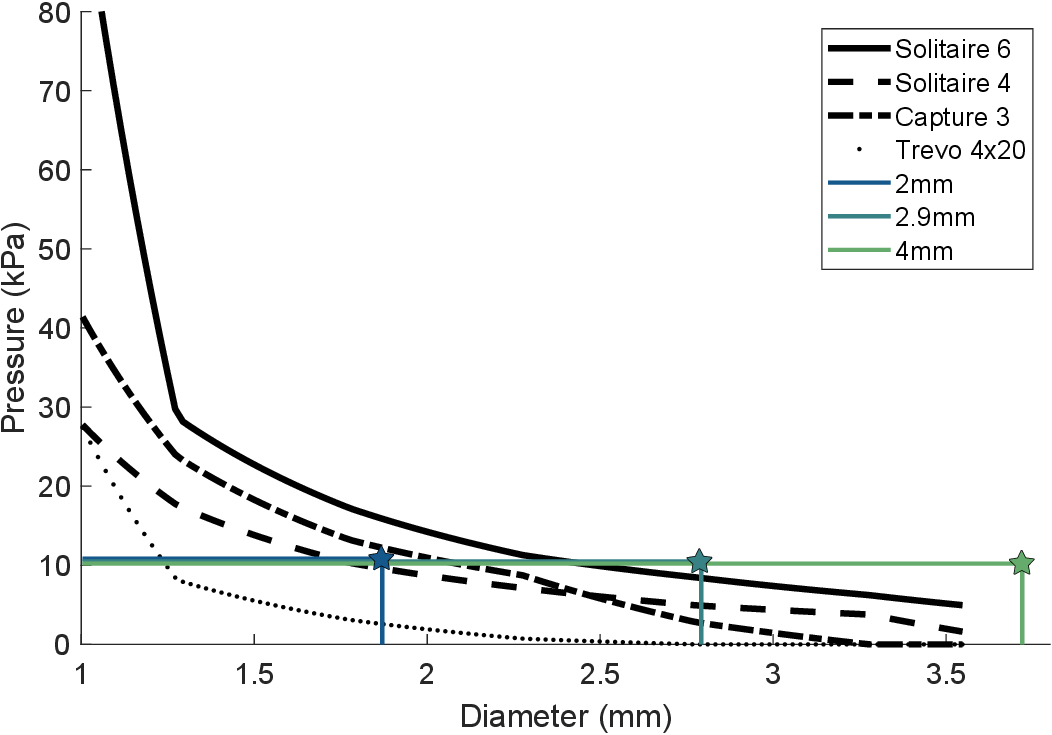}
    \caption{Pressure-diameter curves for four commonly available stents. Overlayed are the pressure and opening diameter requirements to completely damage VSMCs in arteries of three values of original (pre-vasospasm) diameter.}
    \label{fig:stent-pres}
\end{figure}

We compare the pressure predictions for arteries of different physiological (pre-vasospasm in homeostasis) diameters (respectively $2$, $2.9$ and $4$mm) and report these in Table \ref{tbl:fecvs-dmg}. The pressure required for complete damage is about $8.5-11$kPa and the values are very similar between the different arteries, suggesting they are independent of the arterial diameter. However this does not imply equal effectiveness of the stents: indeed the same amount of pressure is required at different diameters and stents apply different pressures at different opening diameters, particularly decreasing pressure as they expand (see Fig, \ref{fig:stent-pres}). To visualise where the maximum required pressure needs to be applied, Figure \ref{fig:stent-pres} shows the pressure-diameter curves for four commonly available stents with overlayed the pressure requirement and corresponding diameter at the time of complete VSMC damage $t_1$ for three arteries of different original diameter (corresponding to the results in Table \ref{tbl:fecvs-dmg_pres}). The point where complete damage would occur is marked with a star: therefore for a given artery-stent pair, the stent would be successful if its curve was above the star mark for the same diameter value at which the star is located. In particular, in the $2mm$ case the required pressure is needed at $1.87mm$, in the $2.9mm$ it must be applied at $2.79mm$, and finally in the $4mm$ case it is needed at $3.72mm$. Therefore according to our model predictions, Solitaire 6 and Capture 3 would be successful in resolving vasospasm in a $2mm$ artery, while all designs would fail in the larger arteries - although with a very small margin for the $2.9$mm case. Considering a margin of error for our model, this is qualitatively consistent with reported clinical cases where the discriminant diameter between likely success and failure of stent-assisted angioplasty is about $3mm$ \cite{Bhogal_Loh_Brouwer_Andersson_Söderman_2017, Bhogal_Paraskevopoulos_Makalanda_2017}.

\begin{table}
    \centering
    \begin{tabular}{c|c|cc|cc|c}
        \textbf{Original diameter} & \textbf{h-element} & $\mathbf{p_{add}(t_0)}$ & $\mathbf{\lambda_{M_{|t_0}}}$ & $\mathbf{p_{add}(t_1)}$ & $\mathbf{\lambda_{M_{|t_1}}}$ & $\mathbf{p_{1D}}$  \\
        \hline
        \multirow{2}*{2mm}   & h1 & 5.21 kPa & 1.30 &  9.20 kPa  & 2.19 & ~8 kPa \\
                             & h8 & 8.47 kPa & 1.30 & 10.83 kPa  & 1.70 & ~8 kPa \\
        \cline{1-7}         
        \multirow{2}*{2.9mm} & h1 & 5.07 kPa & 1.30 &  8.79 kPa  & 2.16 & ~8 kPa \\
                             & h8 & 7.99 kPa & 1.30 & 10.48 kPa  & 1.69 & ~8 kPa \\
        \cline{1-7}  
        \multirow{2}*{4mm}   & h1 & 4.86 kPa & 1.30 &  8.46 kPa  & 2.12 & ~9 kPa \\
                             & h8 & 7.58 kPa & 1.30 & 10.21 kPa  & 1.69 & ~8 kPa \\
    \end{tabular}
    \caption{Comparison of pressures required to initiate damage ($t_0$) and to reach complete damage ($t_1$), and corresponding VSMC stretches, in arteries of different original diameters at $36\%$ stenosis.}
    \label{tbl:fecvs-dmg}
\end{table}

\subsection{Parameter Study: Wall Thickness}
\label{par-study}

\begin{figure}
    \centering
    \includegraphics[scale=0.5]{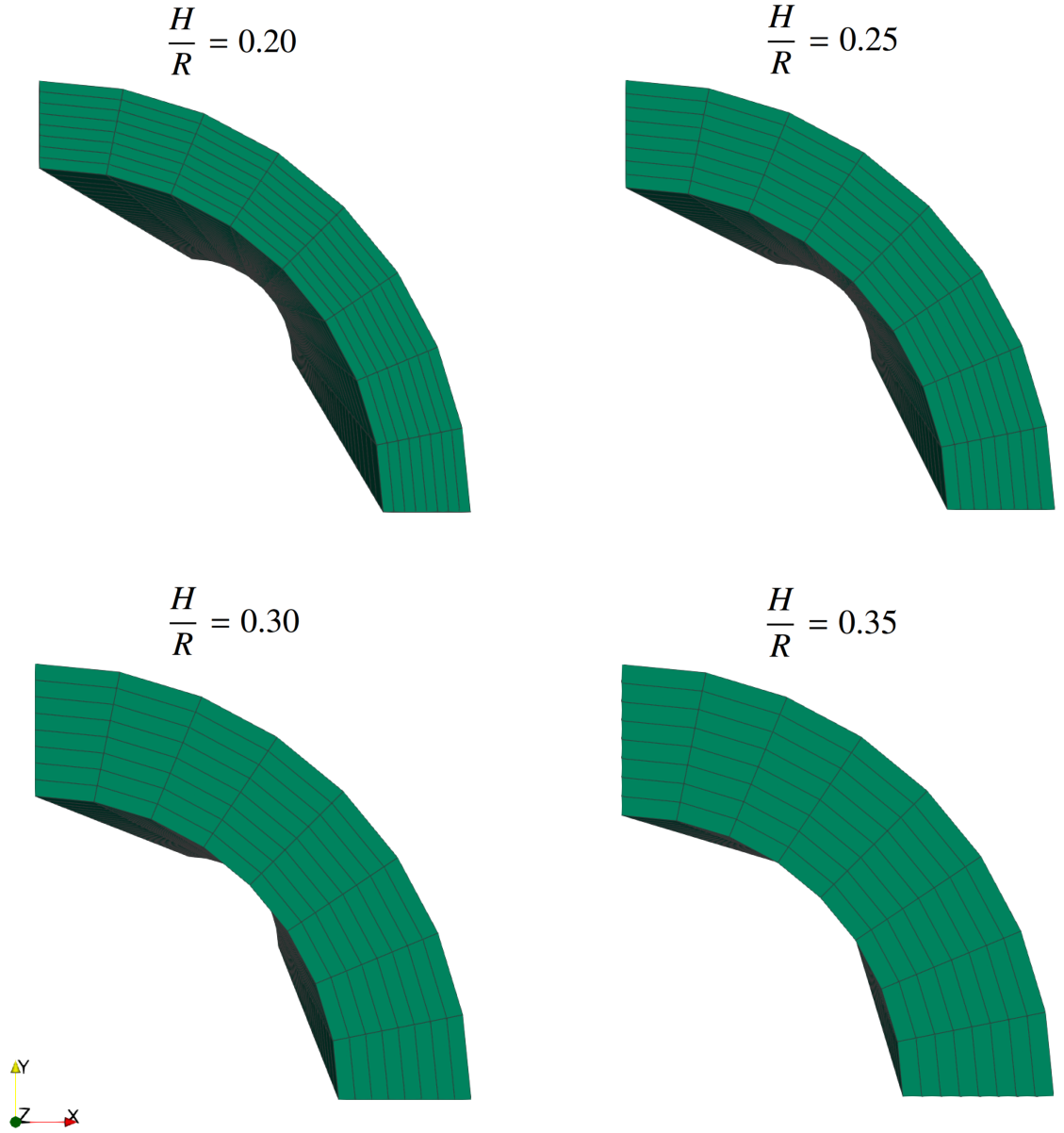}
    \caption{ Axial view of arterial geometry for different levels of arterial thickness-to-radius ratio ($H/R$) considered in the parameter study. }
    \label{fig:hr-panel}
\end{figure}

In this Section we report the results of the parameter study on the effect of the thickness-to-radius ratio $H/R$ on the pressure predictions. We varied the parameter $H/R \in \{0.20, 0.25, 0.30, 0.35\}$. Figure \ref{fig:hr-panel} shows the resulting geometries for the same value of $R=1.45mm$ and Table \ref{tbl:fecvs-dmg-hrparam} reports the predicted pressures at damage onset ($t_0$) and full damage ($t_1$) with the corresponding VSMC stretches. As expected, thicker artery walls result in higher pressure requirements to resolve the disease. Comparing the highest and lowest value of our parameter range, there is a $\sim 2.1$-fold increase in pressure requirement to initiate damage in the innermost layer, $\sim 2.8$-fold increase in the outermost layer, a $\sim 2$-fold increase in pressure requirement to achieve full VSMC damage in the innermost layer, and a $\sim 2.3$-fold increase to achieve full VSMC damage in the outermost layer.

This is a significant difference that might affect the selection of a specific stent model, therefore in the prospect of providing clinical decision support, these results suggest that this parameter requires consideration for patient-specific predictions.

\begin{table}
    \centering
    \begin{tabular}{c|c|cc|cc}
        \textbf{H/R} & \textbf{h-element} & $\mathbf{p_{add}(t_0)}$ & $\mathbf{\lambda_{M_{|t_0}}}$ & $\mathbf{p_{add}(t_1)}$ & $\mathbf{\lambda_{M_{|t_1}}}$  \\
        \hline
        \multirow{2}*{0.20}   & h1 & 1.76 kPa & 1.30 &  3.97 kPa  & 1.94  \\
                              & h8 & 2.26 kPa & 1.30 &  4.68 kPa  & 1.76  \\
        \cline{1-6}         
        \multirow{2}*{0.25}   & h1 & 2.42 kPa & 1.30 &  5.36 kPa  & 1.88  \\
                              & h8 & 3.46 kPa & 1.30 &  6.57 kPa  & 1.69  \\
        \cline{1-6}  
        \multirow{2}*{0.30}   & h1 & 3.15 kPa & 1.30 &  6.79 kPa  & 1.83  \\
                              & h8 & 4.97 kPa & 1.30 &  8.71 kPa  & 1.63  \\
		\cline{1-6}  
        \multirow{2}*{0.35}   & h1 & 3.75 kPa & 1.30 &  7.95 kPa  & 1.79  \\
                              & h8 & 6.38 kPa & 1.30 & 10.67 kPa  & 1.59  \\
    \end{tabular}
    \caption{Comparison of additional pressure requirement and corresponding VSMC stretches to initiate ($t_0$) and complete ($t_1$) damage in the VSMC constituent, for different values of unloaded thickness-to-radius ratio ($H/R$).}
    \label{tbl:fecvs-dmg-hrparam}
\end{table}

\section{Discussion}
\label{sec:discussion}

We have proposed a 3D finite element model of cerebral vasospasm development and mechanical treatment, which to our knowledge is the first of its kind. The model is based on the key assumption that, at the time when mechanical treatment is usually needed, the main factors driving the constriction are VSMC contraction and remodelling, which had been first introduced in the membrane model presented in \cite{Bhogal2019}. Here we propose a 3D finite element model based on a rate-based constrained mixture (rb-CM) framework which allows us to capture the different configurations, growth processes, remodelling laws and damage mechanics of the different biological constituents of the arterial wall. The finite element formulation addresses the limitation of the membrane model by capturing potential hetereogeneities through the wall thickness and in the future will allow the explicit simulation of stent deployment as well as the use of more irregular, realistic arterial geometries.

The model predictions on the magnitude of pressure required to treat the condition are consistent with those exerted by stent-retrievers and about one order of magnitude lower than balloon angioplasty, supporting the validity of stent-assisted angioplasty for the treatment of vasospasm at least in some cases. Stent-retrievers offer significant advantages such as increased safety, easier maneuverability and increased reach of the distal vasculature, so it would be desirable for a clinician to be able to determine whether they are suitable for a specific case and, if so, which specific model. The predictions of the model are consistent with reported clinical observations \cite{Bhogal_Loh_Brouwer_Andersson_Söderman_2017, Bhogal_Paraskevopoulos_Makalanda_2017, Kwon_Lim_Koh_Park_Choi_Kim_Youm_Song_2019, Norby_Young_Siddiq_2019, Badger_Jankowitz_Shaikh_2020, Su_Ali_Pukenas_Favilla_Zanaty_Hasan_Kung_2020, Gupta_Woodward_2022}. The latter suggest that currently available stent-retrievers tend to be effective in arteries of up to about $3mm$ original diameter and the model predictions are consistent with those results. Indeed we obtain that two of the four considered stents would be successful in an artery of $2mm$ original diameter, none would be successful in the $4mm$ case, and the $2.9mm$ is at the threshold of success.

In the parameter study we looked at the effect of the thickness-to-radius ratio on the pressure prediction. Parameter $H/R$ spanned interval $[0.20, 0.35]$ with a step of $0.05$ and resulted in higher pressure requirements in thicker arteries, as would be expected. In particular the case $0.35$ requires about double the amount of pressure compared to case $0.20$ to reach complete VSMC damage in all layers. Although the pre-vasospasm diameter of the affected artery is likely a more important factor to consider in stent selection, the wall thickness is likely to be the second most decisive consideration. The high interpersonal variability of this parameter, which the studies in \cite{Monson_Barbaro_Manley_2008} and \cite{Harteveld_2018} show to depend at least on age and presence of cardiovascular disease, is thus an important parameter to consider when moving towards patient-specific predictions. It is worth noting that changes in thickness may also correspond to morphological changes in the artery wall that would affect its mechanical properties: this is not captured here as the material model remains the same through all simulations. In the future, with the help of advanced imaging techniques and studies on the effect of ageing on the cardiovascular system such as the one described in \cite{valentinMultiLayeredComputationalModel2011}, it will be important to incorporate this aspect in patient-specific predictions. 

The model makes necessary assumptions that require discussion. The most important one is the hypothesis that it is VSMC remodelling that plays a dominant role at the time at which mechanical treatment is usually deemed necessary (about 7-10 days post-SAH, assuming ineffectiveness of pharmacological treatment). In particular it is assumed that, after an acute phase in which the increase of vasoconstrictors and scavengers of vasodilators causes the cells to contract excessively (increase in active response, \cite{Macdonald_Weir_1991, Findlay_Nisar_Darsaut_2016, Baggott_Aagaard-Kienitz_2014}), there is a second phase in which VSMCs remodel, possibly adapting their cytoskeleton and/or focal adhesions in response to this sustained state of contraction. At the time of writing this hypothesis is consistent with all experimental data and observations currently available on cerebral vasospasm \cite{Yamaguchi-Okada_Nishizawa_Koide_Nonaka_2005, Macdonald1995, Matsui1994, Vorkapic1991a, Bevan1987}, and in particular might even explain the conflicting results on collagen remodelling \cite{Smith1985, Nagasawa1982, Yamaguchi-Okada_Nishizawa_Koide_Nonaka_2005}, since it provides a mechanism that causes the observed increase in arterial stiffness and reduction in compliance, without any changes in the collagenous structure. This main hypothesis also provides a possible explanation for the reduced occurrence of vasospasm in the elderly population \cite{pavelkaVasospasmRiskFollowing2023} since the reduced contractile range observed to occur in ageing may result in less significant increases of active contraction and perhaps slower remodelling. This hypothesis is based on observed behaviour of these cells in circumstances other than vasospasm and, given its central role in this model and in the proposed mechanism of action behind stent-based treatment, experimental evidence of VSMC remodelling in vasospasm is warranted. With regards to remodelling, this model also assumes that VSMC stretch has returned to its homeostatic value everywhere in the vessel at the time of treatment. This may not be the case since remodelling is a gradual process and it might be ongoing at the time of treatment, with cell stretch still different from its homeostatic value. If only partial remodelling had occurred at the time of treatment, the model would underestimate the pressure requirements and therefore a thorough analysis should be run on the effect of this parameter on the predictions.

Another limitation of the model is that the damage criterion for VSMCs is based on very limited experimental data and it would benefit from a biologically-informed foundation. The only experimental data available is from \cite{Fischell1990} and the model adopted here is conservative with respect to those results. In the one-dimensional model it was assumed that complete damage occurred at a stretch of about $1.8$ \cite{Fischell1990}. In the finite element model presented here it was not possible to make use of instantaneous damage, but a gradual model had to be adopted: a damage variable was introduced that increases proportionally to the deviation of cell stretch from a defined threshold, and the stress response of the constituent decreases linearly with the damage variable. This model was then calibrated to obtain complete damage at stretches between $1.7$ and $2.2$ as reported in Table \ref{tbl:fecvs-dmg_pres}. This damage criterion has been selected according to the hypothesis that functional failure of the smooth muscle cells is due to either detachment from the extracellular matrix or the breakage of the stress fibers composing their cytoskeleton. However it is possible that other mechanisms would be most suitable and experimental guidance is needed here. Indeed in the selection of an appropriate damage criterion, not only would it be difficult to establish a ``reasonable" parameter space for the one implemented in this model, but the functional form of the criterion itself can be discussed. For example it is possible that cell damage occurs by delamination and the damage propagation law should therefore reflect that behaviour. 

Another assumption made is that damage to VSMCs is sufficient to mechanically resolve the disease. This is consistent with the main hypothesis that VSMC remodelling is a dominating factor, but direct experimental evidence is not currently available. Depending on the time elapsed from SAH it may also be that collagen or other ECM proteins play a no longer negligible role, since at later times collagen growth and remodelling may cause further stiffening of the arterial tissue which would then require higher pressures to resolve the constriction. It would be interesting to study how the inclusion of collagen growth and remodelling might affect the pressure requirement predictions, which would be straight-forward to implement in this framework. Moreover, it is possible that some damage to collagen occurs during treatment via stent-retrievers. In this study we focused on VSMC damage, but future work should include a damage model for collagen and a study of the effect of stent-assisted angioplasty on the collagen network.

In comparison to the one-dimensional model, the predictions on the magnitude of pressure necessary for treatment are a little higher in the thick-walled model: remembering that the $1D$ model simulated a $50\%$ level of stenosis while the thick-walled model simulates $36\%$, the $1D$ model predicted that an additional pressure of about $8kPa$ was necessary to reach the dilatation threshold, while the finite element model presented here predicts that a pressure between $8$ and $11kPa$ is necessary. A higher pressure requirement is expected and reasonable for the thick-walled model compared to the membrane, although a direct comparison of numerical values is not entirely appropriate: indeed the $1D$ model considers the experimental measurements of the outward pressure applied by the stents and the fact that this pressure is higher when the stent is ``compressed" just before deployment and gradually decreases as the stent expands; on the contrary the finite element implementation of this model uses an \textit{increasing} internal pressure which is less accurate in representing the behaviour of a stent. As a consequence, the $1D$ model predicts an additional pressure of $8kPa$ \textit{at} the dilatation threshold, but the stent must have been able to apply higher pressures before reaching that threshold or it would not have been able to dilate the artery up until that level of stretch. 

Another limitation is the simulation of treatment, which here consists of a simple increase in internal pressure. In the one-dimensional model by Bhogal et al. \cite{Bhogal2019} it had been possible to use the chronic outward force data provided by stent manufacturers, which decreases as the stent expands. This could be a non-negligible difference since the amount of force exerted by a stent decreases with expansion, while in this model it linearly increases. Moreover the increase in internal pressure in the presented model is applied uniformly across the luminal surface, whereas during stent deployment a larger force would be applied to the constricted area of the vessel compared to the unaffected area. In the future it is recommended to develop an explicit finite element model of stent deployment inside a thick-walled vessel, incorporating surface contact with the vessel wall and a mesh that explicitly captures the geometric design and material properties of the specific stent model. We believe this would also overcome the numerical instabilities we encounter in some simulations and would make the treatment predictions more accurate and realistic. 

In order to progress the model towards a clinical decision support tool, several improvements are recommended. The idealised cylindrical geometry of the artery should be changed to a more realistic irregular shape, ideally informed by real geometries. The stent geometry should also be represented explicitly since the mesh density and material properties might affect the application of pressure on the arterial wall and in particular the propagation of cell damage across the wall thickness (\cite{Geith2020}). Finally the damage propagation model should be validated or changed accordingly to experimental studies. The arterial wall should be modelled as a two-layered structure to represent the different mechanical properties of the medial and adventitial layers. In the future it would also be of interest to include the fluid dynamics of blood flow within the artery and its interaction with the vessel wall. 

Previous models of vasospasm have focused either entirely on the cerebral haemodynamics \cite{Lodi_Ursino_1999} or on a 2D coupling between the haemodynamics and the growth and remodelling processes within the arterial wall \cite{Humphrey_Baek_Niklason_2007, Baek2007}, which was able to capture the potentially self-resolving nature of the disease. In the presented work we adopted a very similar approach to the latter: we also used a constrained mixture model, but made use of a more computationally efficient rate-based approach (rb-CMM); we considered elastin, collagen and vascular smooth muscle cells as structural constituents and used 3D-analogues of the same material models for elastin and VSMCs; finally we initiated vasospasm as an increase in the active stress of VSMCs. In contrast, we omitted the coupling to the haemodynamics in order to focus on the time at which mechanical treatment is required. We incorporated the hypothesis that (at the time of mechanical treatment) the constriction is driven by VSMC remodelling, which was first presented in \cite{Bhogal2019}, and implemented a gradual strain-based damage model for VSMCs. 

In this work we included the aspects that we deemed more crucial for the goal of studying requirements for mechanical treatment, but in the future it would be of interest to develop a 3D fluid-structure interaction model where arterial biomechanics and mechanobiology are coupled to blood flow dynamics, for example adopting the framework used in \cite{Tricerri_Dedè_2015, Teixeira_Neufeld_2020} or \cite{Moerman_Konduri_2022}, and considering the dependence of VSMC stress response on arterial stress distributions \cite{Rachev_Hayashi_1999}. This would further improve understanding on the development and progression of the disease and in the future could help predict how well a patient would recover after treatment since the restoration of blood flow through the vessel would affect brain tissue perfusion.

The rate-based constrained mixture framework described in this work is a state-of-the-art code base for soft tissue growth and remodelling simulations. It was first proposed in membrane models of fusiform and saccular aneurysms in \cite{watton_mathematical_2004, watton2009} where the concept of recruitment stretch was introduced and suggested as a key remodelling mechanism to explain aneurysm evolution. It was adapted to a 3D framework in \cite{Schmid_Watton_Maurer_Wimmer_Winkler_Wang_Röhrle_Itskov_2010} and then extended to include isotropic volumetric growth \cite{eriksson2014} and finally anisotropic volumetric growth \cite{grytsan2017}. The framework adopts a constrained mixture approach which allows the representation of distinct properties, configurations and growth and remodelling laws of the different tissue constituents. It already includes ground matrix, elastin, collagen and VSMCs but is easily adapted to include other materials or exclude existing ones. Common material models are readily available, including one for collagen with undulation distribution and one for VSMCs with an active stress response, which were presented in this work. The rate-based approach for growth and remodelling laws avoids numerical integration and keeps computational costs low (a simulation with 3072 nodes and 6500 time steps took 4 hours and 18 minutes on a single 12th Gen Intel(R) Core(TM) i7-12700H 2.30 GHz). It is possible to model anisotropic volumetric growth and coupling to computational fluid dynamics solvers for fluid-solid-growth simulations is also straight-forward \cite{grytsan2015}. Finally it includes a damage model that can be based on a custom criterion, where a strain-based one has been exemplified in the presented work. Due to its versatility, the code base represents a flexible and efficient framework that can be easily adapted to modelling other soft tissue conditions, for example aneurysms and bladder outlet obstruction \cite{Teixeira_Neufeld_2020, cheng_constrained_2022}.

\section{Conclusion}
\label{sec:conclusions}

We have developed a novel rate-based constrained mixture model accounting for vascular smooth muscle cells active stress, remodelling and damage, and integrated it into a finite element framework which we applied to simulate vasospasm development and treatment. The model predictions regarding pressure requirements for mechanical treatment are consistent with reported clinical observations, namely that the magnitude of pressure required is closer to that of stents than of balloon angioplasty, and that currently available designs are in general successful in arteries of up to about $3mm$ original diameter, but tend to fail in larger ones. The extended framework provides the foundation for more refined simulations with realistic geometries and explicitly resolved stent deployment, which with future work may lead to clinical decision support tools or computational aids for the design of stents designed specifically for treatment of cerebral vasospasm. 

\section*{Statements of Ethical Approval}\label{sec:EthicalApproval}
The authors do not need ethical approval.

\section*{Author Contributions}
\textbf{Giulia Pederzani}: Conceptualization, Formal Analysis, Methodology, Software, Visualization, Writing - Original draft; \textbf{Andrii Grytsan}: Software, Writing - Review \& Editing; \textbf{Anne Robertson}: Supervision (supporting), Writing - Review \& Editing; \textbf{Alfons Hoekstra}: Supervision (supporting), Funding Acquisition; \textbf{Paul Watton}: Supervision (lead), Methodology, Project Administration, Funding Acquisition, Writing - Review \& Editing.

\section*{Funding}\label{sec:Funding}
GP acknowledges a PhD scholarship provided by the Department of Computer Science, University of Sheffield, and the EDITH coordination and support action funded by the Digital Europe program of the European Commission (grant n. 101083771). PNW acknowledges partial support towards this work from UK EPSRC (EP/N014642/1 and EP/T017899/1).

\section*{Declaration of Competing Interest}\label{sec:CompetingInterest}
The authors declare no competing interests.





\bibliographystyle{elsarticle-num}
\bibliography{biblio.bib}







\end{document}